\documentclass{revtex4}
\usepackage{amssymb}
\usepackage{graphicx}
\begin{document}
\title{ A spectroscopy approach to measure the gravitational mass of antihydrogen }
\author{A. Yu. Voronin$^1$, V.V. Nesvizhevsky$^2$, G. Dufour$^3$, P. Debu$^4$, A. Lambrecht$^3$, S. Reynaud$^3$, O.D. Dalkarov$^1$, E.A. Kupriyanova$^1$, P.Froelich$^5$}
\affiliation{ $^1$ P.N. Lebedev Physical Institute, 53 Leninsky
prospect, 117924 Moscow, Russia.
\\
$^2$ Institut Max von Laue - Paul Langevin (ILL), 6 rue Jules Horowitz, F-38042, Grenoble, France.
\\
$^3$ Laboratoire Kastler-Brossel, CNRS, ENS, UPMC, Campus Jussieu, 75252, Paris, France.
\\
$^4$ Institut de Recherche sur les lois Fondamentales de l'Univers, CEA-Saclay, 91191, Gif-sur-Yvette, France.
\\
$^5$ Department of Quantum Chemistry, Uppsala University, Box 518,
SE-75120 Uppsala, Sweden.
}

\begin{abstract}

We study a method to induce resonant transitions between antihydrogen ($\bar{H}$) quantum states above a material surface in the gravitational field of the Earth. The method consists of applying a gradient of magnetic field, which is temporally oscillating with the frequency equal to a frequency of transition between gravitational states of antihydrogen. A corresponding resonant change in the spatial density of antihydrogen atoms could be measured as a function of the frequency of applied field. We estimate an accuracy of measuring antihydrogen gravitational states spacing and show how a value of the gravitational mass of the $\bar{H}$ atom could be deduced from such a measurement. We also demonstrate that a method of induced transitions could be combined with a free-fall-time measurement in order to further improve the precision.

\end{abstract}

\maketitle

\section{Introduction}
Precision tests of the Equivalence Principle (EP) in various physical phenomena is of fundamental interest. This statement is especially valid for testing the EP in case of a quantum motion of antiatoms. Detailed studies of gravitational properties of antimatter are planned to be performed by most groups involved in experiments with antihydrogen ($\bar{H}$) \cite{Alpha,AlphaGrav, Yam, gabr10, cesa05, Aegis1}.

Here we discuss a possibility of exploring gravitational states of $\bar{H}$ \cite{GravStates} using potentially very precise spectroscopic methods. We study the behaviour of $\bar{H}$ bounded in the gravitational field near a material surface under the influence of an alternating magnetic field with a frequency adjusted to induce resonance transitions between lowest gravitational states. This approach allows an access to the gravitational energy level spacing and thus to the gravitational mass of $\bar{H}$.

\section{ Gravitationally bound Antihydrogen in an alternating magnetic field }
%In the following we will study transitions between  gravitational states of $\bar{H}$ center of mass (c.m), which are induced by an alternating magnetic field.
In this section we will study a motion of an $\bar{H}$ atom, localized in a gravitational state near a horizontal plane mirror, under the influence of an alternating magnetic field.

The interaction of a magnetic field with \emph{a moving through the field} ground state $H$ atom \cite{LambH,GorkH,LozH} is dominated by the interaction of an average magnetic moment of the atom \cite{LL} in a given hyperfine state with the magnetic field. We are going to focus on an alternating magnetic field with a gradient in the vertical direction. This condition is needed for coupling the field and the center of mass (c.m) $\bar{H}$ motion in the gravitational field of the Earth. It allows one to induce resonant transitions between quantum gravitational states of $\bar{H}$ \cite{GravStates}. Such states have similar properties with those discovered for neutrons \cite{Nature1,nesv00,nesv03,NVP,EPJC,NeutrWaveGuide}. In particular they are characterised by the following characteristic energy ($\varepsilon_n$) and spatial ($H_n$) scales:
\begin{eqnarray}\label{eGrav}
&\varepsilon_0&=\sqrt[3]{\frac{\hbar^2M^2g^2}{2m}},\\
&l_0&=\sqrt[3]{\frac{\hbar^{2}}{2mMg}},\\
&\varepsilon_n&=\varepsilon_0 \left(\lambda_n+\Delta \right),\\
&H_n&=l_0 \lambda_n, \label {Hn} \\
&\mathop{\rm Ai}(&-\lambda_n)=0.
\end{eqnarray}
 Here $M$ is the gravitational mass of $\bar{H}$, $m$ is the inertial mass of $\bar{H}$ (we distinguish between $M$ and $m$ in view of discussing EP tests), $g$ is the gravitational field intensity near the Earth surface, $\mathop{\rm Ai}(x)$ is an Airy function \cite{abra72}, $\Delta\simeq-i0.005$ is a universal complex shift of gravitational levels due to the interaction of $\bar{H}$ with a material (perfectly conducting) wall \cite{GravStates}. All states acquire equal width, which is a function of a material surface substance $\Gamma=2\sqrt[3]{\frac{\hbar^2M^2g^2}{2m}}|\Delta|$. This width corresponds to the lifetime of $0.1$ s in case of a perfectly conducting surface and is twice longer for silica \cite{voro05,QRslabs,QRnanopor} for instance.

   % The problem of an atom motion in nonhomogeneous alternating magnetic field is complicated by coupling of the center of mass (c.m.) motion with interatomic degrees of freedom including spin.
    %However, we will show that for the field intensities and characteristic frequencies of interest the adiabatic approximation is justified, which means that internal state of an atom "follows" at any point in space and at any %instant of time the value of magnetic field $\vec{B}(\vec{r},t)$.

   % The problem of separation of the center of mass motion and relative motion in case of hydrogen atom  in homogeneous magnetic field  was solved in %\cite{Lamb,Dzyal}. The problem we are interested in here is more complicated, % namely antihydrogen atom bounces in the gravitational field of the %Earth above material surface superposed with \emph{nonhomogeneous} time-varying magnetic field $\vec{B}$.
    We will consider the magnetic field in the following form:
% \begin{equation}\label{Magn}
%\vec{B}(z,x,t)=\left(B_0+ B_z(z,x,t)\right) \vec{e}_z+B_x(z,x,t) \vec{e}_x
%\end{equation}
\begin{equation}\label{Magn1}
\vec{B}(z,x,t)=B_0 \vec{e}_z+ \beta \cos(\omega t) \left(z \vec{e}_z-x \vec{e}_x \right).
\end{equation}

Here $B_0$ is the amplitude of a constant, vertically aligned, component of magnetic field, $\beta$ is the value of magnetic field gradient, $z$ is a distance measured in the vertical direction, $x$ is a distance measured in the horizontal direction, parallel to the surface of a mirror.

A time-varying magnetic field (\ref{Magn1}) is accompanied with an electric field ($[\vec{\nabla} \vec{E}]=-\frac{1}{c} \partial \vec{B}/\partial t$). However, for the velocities of ultracold atoms, corresponding interaction terms are small and thus will be omitted.

An inhomogeneous magnetic field couples the spin and the spatial degrees of freedom. A $\bar{H}$ wave-function is described in this case using a four-component column (in a nonrelativistic treatise) in the spin space, each component being a function of the c.m. coordinate $\vec{R}$, relative $\bar{p}-\bar{e}$ coordinate $\vec{\rho}$ and time $t$.
The corresponding Schr\"{o}dinger equation is:
\begin{equation} \label{Schr}
i \hbar\frac{\partial \Phi_{\alpha}(\vec{R},\vec{\rho},t)}{\partial t}=\sum_{\alpha'}\left[ -\frac{\hbar^2}{2m}\Delta_R+Mgz+V_{CP}(z) +\widehat{H}_{in}+\widehat{H}_m \right]_{\alpha, \alpha'} \Phi_{\alpha'}(\vec{R},\vec{\rho},t).
\end{equation}
A subscript $\alpha$ in this equation indicates one of four spin states of the $\bar{p}-\bar{e}$ system.
The meaning of the interaction terms is the following.

$V_{CP}(z)$ is an atom-mirror interaction potential, which turns into the van der Waals/ Casimir-Polder potential at an asymptotic atom-mirror distance (see \cite{voro05,voro05l} and references therein).
$\widehat{H}_{in}$ is the Hamiltonian of the internal motion, which includes the hyperfine interaction.
\begin{equation}\label{Hc}
\widehat{H}_{in}=-\frac{\hbar^2}{2 \mu}\Delta_\rho-e^2/\rho+\frac{\alpha_{HF}}{2}\left(\hat{F}^2-3/2\right).
\end{equation}
Here $\mu=m_1m_2/m$, $m_1$ is the antiproton mass, $m_2$ is the positron mass, $m=m_1+m_2$, $\alpha_{HF}$ is the hyperfine constant, $\hat{F}$ is the operator of the total spin of the antiproton and the positron. We will treat only $\bar{H}$ atoms in a $1S$-state (below we will show that the excitation of other states in the studied process is improbable). The term $\frac{\alpha_{HF}}{2}\left(\hat{F}^2-3/2\right)$ is a model operator, which effectively accounts for the hyperfine interaction and reproduces the hyperfine energy splitting correctly.

The term $\widehat{H}_m$ describes the field-magnetic moment interaction:
\begin{equation}\label{Hm}
\widehat{H}_m=-2\vec{B}(z,x,t)\left( \mu_{\bar{e}}\hat{s}_{\bar{e}}\times \hat{I}_{\bar{p}}+ \mu_{\bar{p}}\hat{s}_{\bar{p}}\times \hat{I}_{\bar{e}}\right).
\end{equation}
Here $\mu_{\bar{e}}$ and $\mu_{\bar{p}}$ are magnetic moments of the positron and the antiproton respectively, $\hat{s}_{\bar{e}}$, $\hat{s}_{\bar{p}}$ is a spin operator, acting on spin variables of positron (antiproton), $\hat{I}_{\bar{e}}$, $\hat{I}_{\bar{p}}$ is a corresponding identity operator.

As far as the field $\vec{B}(z,x,t)$ changes in space and in time, this term couples the spin and the c.m. motion.

We will assume that in typical conditions of a spectroscopy experiment the $\bar{H}$ velocity component $v$ parallel to the mirror surface (directed along $x$-axis) is of the order of a few $m/s$ and is much larger than a typical vertical velocity in lowest gravitational states (which is of the order of $cm/s$). We will treat the motion in a frame moving with the velocity $v$ of the $\bar{H}$ atom along the mirror surface. Thus we are going to consider the x-component motion as a classical motion with a given velocity $v$, and we will substitute a $x$-dependence by a $t$-dependence. We will also assume that $B_0\gg \beta L$, where $L\sim 30$ cm is a typical size of an experimental installation of interest. This condition is needed for "freezing" the magnetic moment of an atom along the vertical direction; it provides the maximum transition probability.

We will be interested in the weak field case, such that the Zeeman splitting is much smaller than the hyperfine level spacing $\mu_B B_0\ll \alpha_{HF}$. The hierarchy of all mentioned above interaction terms could be formulated as follows:
\begin{equation}\label{hierar}
m_2e^2/\hbar^2 \gg \alpha_{HF}\gg \mu_{\bar{e}} |B_0 |\gg \varepsilon_n,
\end{equation}
and thus it justifies the use of the adiabatic expansion for solving Eq.(\ref{Schr}); it is based on the fact that an internal state of an $\bar{H}$ atom follows adiabatically the spatial and temporal variations of an external magnetic field. Neglecting non-adiabatic couplings, an equation system for the amplitude $C_n(t)$ of a gravitational state $g_n(z)$ has the form:
\begin{equation}\label{Adiab}
i \hbar \frac{d C_{n}(t)}{dt}= \sum_{k} C_{k}(t) V(t)_{n,k}\exp \left(-i\omega_{n k} t\right ).
\end{equation}
The transition frequency $\omega_{n k}=(\varepsilon_k-\varepsilon_n)/\hbar$ is determined by the gravitational energy level spacing. This fact is used in the proposed approach to access the gravitational level spacing by means of scanning the applied field frequency, as will be explained in the following.

Within this formalism the role of the coupling potential $V(z,t)$ is played by the energy of an atom in a fixed hyperfine state thought of as a function of (slowly varying) distance $z$ and time $t$.
\begin{equation}
V(t)_{n,k}=\int_0^\infty g_n(z) g_{k}(z) E(t,z)dz.\\
\end{equation}
Here $g_n(z)$ is the gravitational state wave-function, which is given using the Airy function \cite{GravStates}.

The energy $E(z,t)$ is the eigenvalue of the internal and magnetic interactions $\widehat{H}_{in}+\widehat{H}_m$, where the c.m. coordinate $\vec{R}$ and time $t$ are treated as slow-changing parameters.
Corresponding expressions for the eigen-energies of a $1S$ manifold are:

\begin{eqnarray}\label{Ea}
E_{a,c}&=&E_{1s}-\frac{\alpha_{HF}}{4}\mp\frac{1}{2}\sqrt{\alpha_{HF}^2+|(\mu_B-\mu_{\bar{p}})B(z,t)|^2},  \\ \label{Eb}
E_{b,d}&=&E_{1s}+\frac{\alpha_{HF}}{4}\mp \frac{1}{2}|(\mu_B+\mu_{\bar{p}})B(z,t)|.
\end{eqnarray}

Subscripts $a,b,c,d$ are standard notations for hyperfine states of a $1S$ manifold in a magnetic field.
The presence of a constant field $B_0$ produces the Zeeman splitting between states $b$ and $d$. As far as the energy of states $b,d$ depends on magnetic field linearly, while for states $a,c$ it depends quadratically, only transition between $b,d$ states take place in case of a weak field. In the following we will consider only transitions between gravitational states in a $1S( b,d)$ manifold.

A qualitative behavior of the transition probability is given in the Rabbi formula, which can be deduced by means of neglecting the high frequency terms compared to the resonance couplings of only two states, initial $i$ and final $f$, in case the field frequency $\omega$ is close to the transition frequency $\omega_{if}=(E_f-E_i)/\hbar$:
\begin{equation}\label{Rabbi}
P=\frac{1}{2}\frac{(V_{if})^2}{(V_{if})^2+\hbar^2(\omega-\omega_{if})^2}\sin^2\left(\frac{\sqrt{(V_{if})^2+\hbar^2(\omega-\omega_{if})^2}}{2\hbar}t\right)\exp(-\Gamma t).
\end{equation}
 The factor $1/2$ appears in front of the right-hand side of the above expression due to the fact that only two $(b,d)$ of four hyperfine states participate in the magnetically induced transitions.

It is important that the transition frequencies $\omega_{if}$ do not depend on the antiatom-surface interaction up to the second order in the spliting $\Delta$. This is a consequence of the already mentioned fact that all energies of gravitational states acquire equal (complex) shift due to the interaction with a material surface.

\begin{figure}
 \centering
\includegraphics[width=100mm]{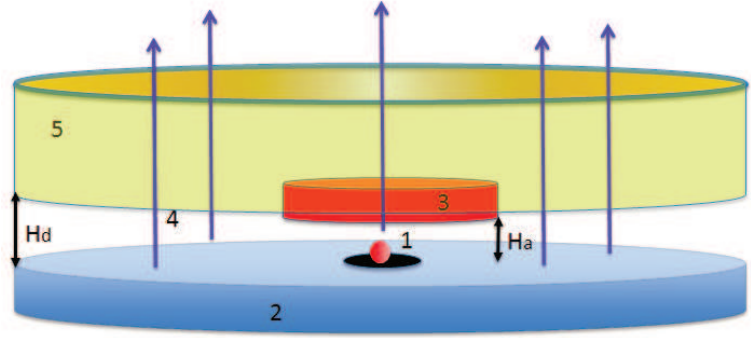}
\caption{A sketch of the principle scheme of an experiment on magnetically induced resonant transitions between $\bar{H}$ gravitational states. 1 - a source of ultracold antihydrogen, 2 - a mirror, 3 - an absorber, 4 - a magnetic field, 5 - a detector.}\label{FigSketch}
\end{figure}

A resonant spectroscopy of $\bar{H}$ gravitational states could consist of observing $\bar{H}$ atoms localized in the gravitational field above a material surface at a certain height as a function of the applied magnetic field frequency.
 A "flow-through type" experiment, analogous to the one, discussed for the spectroscopy of neutron gravitational states \cite{ResGranit}, includes three main steps. A sketch of a principle scheme of an experiment, proposed in \cite{Shaping}, is shown in Fig.\ref{FigSketch}.

%%%%%%%%%%%%%%%%%%%%%%%%%%%%%%%%%%%%%%%%%%%%%%%%%%%%%%%%%%%%%%%%%%%%%%%%
%%%%%%%%%%%%%%%%%%%%%%%%%%%%%%%%%%%%%%%%%%%%%%%%%%%%%%%%%%%%%%%%%%%%%%%%%%%

First, an atom of $\bar{H}$ is shaped in a ground gravitational state. This is achieved by means of passing $\bar{H}$ through a slit, formed by a mirror and an absorber, which is placed above the mirror at a given height $H_a$. The mirror and the absorber form a waveguide with a state-dependent transmission \cite{NeutrWaveGuide}. The choice of $H_a=H_1\simeq 13.6$ $\mu m$ provides that only $\bar{H}$ atoms in the ground gravitational state pass through the slit.

Second, $\bar{H}$ atoms are affected by an alternating magnetic field (\ref{Magn1}) while they are moving parallel to the mirror. An excited gravitational state is resonantly populated.

Third, the number of $\bar{H}$ atoms in an excited state is measured by means of counting the annihilation events in a detector, which is placed at a height $H_d$ above the mirror. The value of $H_d$ is chosen to be larger than the spatial size of the gravitational ground state and smaller than the spatial size of the final state (\ref{Hn}), $ H_1\ll H_d<H_f$, so that the ground state atoms pass through, while atoms in the excited state are detected.

%Velocity of ultracold $\bar{H}$ as estimated in \cite{Gbar} is of order of $1$ m/s

\begin{figure}
 \centering
\includegraphics[width=100mm]{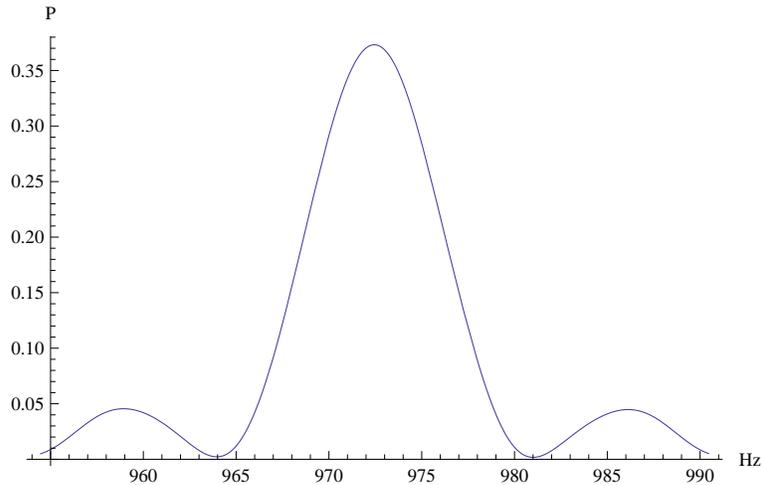}
\caption{The transition probability as a function of the magnetic field frequency for the transition from the ground state to $6$-th gravitational state.}\label{FigTrans}
\end{figure}

We present a simulation of the number of detected annihilation events as a function of the field frequency in Fig.\ref{FigTrans} for the transition from the ground to the $6$-th excited state, based on a numerical solution of the equation system Eq.(\ref{Adiab}). The corresponding resonance transition frequency is $\omega=972.46$ Hz. The value of the field gradient, optimized to obtain the maximum probability of $1\rightarrow 6$ transition during the time of flight $t_{fl}=\tau=0.1$ s, turned to be equal $\beta=27.2$ Gs/m, the corresponding guiding field value, which guarantees the adiabaticity of the magnetic moment motion, is $B_0=30$ Gs.
%%%%%%%%%%%%%%%%%%%%%%%%%%%%%%%%%%%%%%%%%%%%%%%%%%%%%%%%%%%%%%%%%%%%%%%%
%%%%%%%%%%%%%%%%%%%%%%%%%%%%%%%%%%%%%%%%%%%%%%%%%%%%%%%%%%%%%%%%%%%%%%%%%%%

It follows from (\ref{eGrav}) that the $\bar{H}$ gravitational mass could be deduced from the measured transition frequency $\omega_{nk}$ as follows:
\begin{equation}
M=\sqrt{\frac{2m\hbar \omega_{nk}^3 }{g^2(\lambda_k-\lambda_n)^3}}.
\end{equation}

Let us mention that $g$ in the above formula means the \emph{gravitational field intensity} near the Earth surface, a value which characterizes properties of the field and is assumed to be known with a high precision. At the same time all the information about gravitational properties of $\bar{H}$ is included in the gravitational mass $M$.

Equality of the gravitational mass $M$ and the inertial mass $m$, imposed by the Equivalence principle, results in the following expression:
\begin{equation}
M=\frac{2\hbar \omega_{nk}^3 }{g^2(\lambda_k-\lambda_n)^3}.
\end{equation}

Assuming that the spectral line width is determined by the lifetime $\tau\approx 0.1$ s of gravitational states, we estimate that the gravitational mass $M$ can be deduced with the relative accuracy $\epsilon_M\sim 10^{-3}$ for $100$ annihilation events for the transition to the $6$-th state.

\section {Free fall of a superposition of gravitational states.}

\begin{figure}
 \centering
\includegraphics[width=100mm]{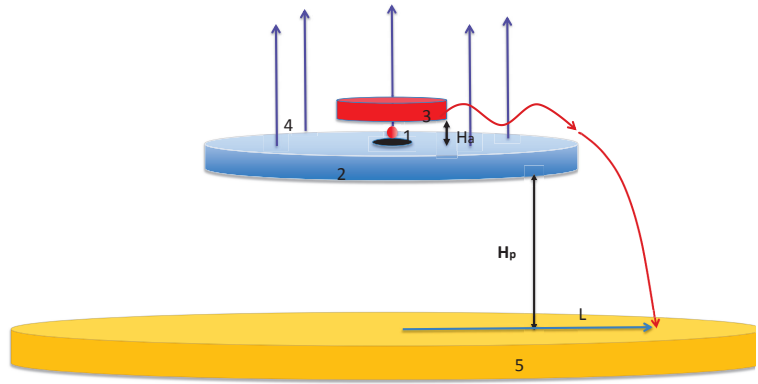}
\caption{A sketch of the principle scheme of an experiment on free fall of a superposition of $\bar{H}$ gravitational states. 1 - a source of ultracold antihydrogen, 2 - a mirror, 3 - an absorber, 4 - a magnetic field, 5 - a detector plate.}\label{FigSketch4}
\end{figure}

An additional option in measuring the gravitational properties of antiatoms consists of measuring the time-distribution of free-fall events. Such a possibility assumes an experimental setup, in which the moment of an antiatom released from the trap is well defined. This property is achieved in the Gbar \cite{Yam} experimental scheme by means of using short laser pulses for photo-detaching the positrons from pre-cooled and trapped $\bar{H}^{+}$ ions. Such a pulse "releases" neutral $\bar{H}$ from an electrostatic trap, and it could be used as a "time zero" signal. It was shown in \cite{Shaping} that the highest accuracy could be achieved in case of the velocity of released atoms being "shaped" by means of a mirror-absorber setup.  After passing through the mirror-absorber shaping device $\bar{H}$ falls down from the edge of the mirror to the horizontal detection plate, positioned at a height $H_p=-30$ cm below the mirror, as shown in Fig.\ref{FigSketch4}.
%%%%%%%%%%%%%%%%%%%%%%%%%%%%%%%%%%%%%%%%%%%%%%%%%%%%%%%%%%%%%%%%%%%%%%%%
%%%%%%%%%%%%%%%%%%%%%%%%%%%%%%%%%%%%%%%%%%%%%%%%%%%%%%%%%%%%%%%%%%%%%%%%%%%
Now the detector plate is a position-sensitive detector, which allows one to measure the horizontal velocity of antiatoms via the relation $V=L_d/T$, where $L_d$ is a radial position of an annihilation event on the detection plate, $T$ is a time of flight starting from the moment of release. The value of horizontal velocity enables one to calculate the time spent by the atom on the mirror before falling down $t=L/V=LT/L_d$. With this correction one can find the time distribution of free-fall events. We are going to demonstrate that such a distribution is closely related to the momentum distribution $F_0(p)$ in an "initial"  wave-packet of $\bar{H}$ atoms taken at the moment when such a wave-packet is at the edge of the mirror. The wave-packet free-fall evolution is determined using the free-fall propagator $G(p,p',t)$:
\begin{eqnarray}\label{FreeFallEv}
\Psi(z,t)=\frac{1}{\sqrt{2\pi \hbar}}\int_{-\infty}^{\infty} \int_{-\infty}^{\infty} G(p,p',t)F_0(p')\exp\left(ipz/\hbar\right)) dp dp',\\ \label{GFF}
G(p,p',t)=\exp\left(-\frac{it}{2m\hbar}(p^2-Mgpt+M^2g^2t^2/3)\right) \delta(p-Mgt-p').
\end{eqnarray}

For sufficiently large times of free fall $t\gg\hbar/\varepsilon_0$ the integral (\ref{FreeFallEv}) could be estimated using the stationary phase method, which gives:
\begin{equation}\label{FFstat}
\Psi(z,t)\simeq \sqrt{\frac{m}{t}}\exp\left(\frac{imz^2}{2t\hbar}-\frac{it^3M^2g^2}{2m\hbar}\right) F_0(p_0-Mgt),
\end{equation}
where $p_0=(z+gt^2/2)m/t$ is a stationary phase point.

The above expression maps the initial momentum distribution $F_0(p)$ into the time- or position- distribution for large times $t$. In particular it means that a time distribution $P$ of free-fall events of a wave-packet, made of a coherent superposition of a few gravitational states (or only one gravitational state) is determined by a velocity distribution in the initial wave-packet, taken at the moment when this wave-packet started the free fall:
\begin{equation}\label{pt}
P=\frac{m}{t}|F_0\left(Mg(t-\sqrt{2mH_p/(Mg)})\right)|^2.
\end{equation}
Here $H_p$ is a height of fall.

The distribution of initial velocities in the wave packet in the form of a superposition of two gravitational states ($g_1$ and $g_6$) is illustrated in Fig. 4. The corresponding distribution of free-fall-times is presented in Fig. 5.
%%%%%%%%%%%%%%%%%%%%%%%%%%%%%%%%%%%%%%%%%%%%%%%%%%%%%%%%%%%%%%%%%%%%%%%%
\begin{figure}
 \centering
\includegraphics[width=80mm]{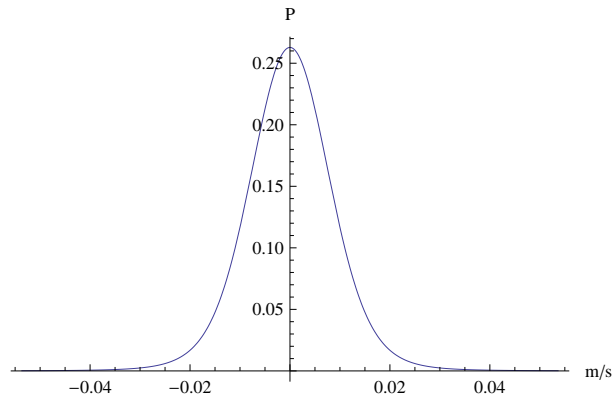}
\caption{The velocity distribution in a ground gravitational state.}\label{FigVelocity}
\end{figure}
%%%%%%%%%%%%%%%%%%%%%%%%%%%%%%%%%%%%%%%%%%%%%%%%%%%%%%%%%%%%%%%%%%%%%%%%%%%
\begin{figure}
 \centering
\includegraphics[width=80mm]{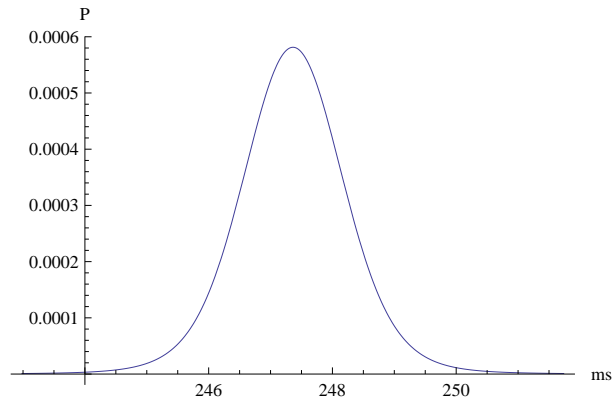}
\caption{The distribution of the time of free-fall events of a wave-packet, comprised of a ground gravitational state. The height of free fall $H_p=30$ cm.}\label{FigTime}
\end{figure}
%%%%%%%%%%%%%%%%%%%%%%%%%%%%%%%%%%%%%%%%%%%%%%%%%%%%%%%%%%%%%%%%%%%%%%%%%%%

The corresponding position of a maximum in the time-of-fall distribution is equal $t_0= 247.4$ ms, that corresponds to a free fall time $t_0=\sqrt{2mH_p/(Mg)}$ from the height of $30$ cm. The width of the distribution is $\delta=1.9$ ms, that gives an accuracy of measuring $M/m$ with $1000$ annihilation events of the order of $10^{-4}$.

A resonant transition induced by a magnetic field could be used for further improving the accuracy of measurements and for gaining additional physical information. In case of $\bar{H}$ passing through a region of an oscillating magnetic field the superposition of two resonantly coupled gravitational states could be produced with a desired relative phase. The corresponding superposition wave-function in case of a resonant field is given in two-state approximation in the following expression:
  \begin{equation}
  \Psi(z,t)\sim cos (|V_{if}|t/\hbar )g_i(z)-i \sin(|V_{if}|t/\hbar ) \exp\left(-i(E_f-E_i)t/\hbar\right ) g_f(z)
  \end{equation}

\begin{figure}
 \centering
\includegraphics[width=80mm]{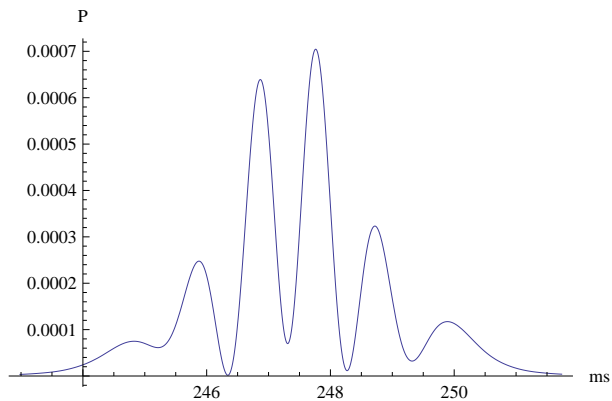}
\caption{The distribution of times of free-fall events for a wave-packet, comprised of a superposition of first and sixth  gravitational states. The height of free fall is $H_p=30$ cm.}\label{FigFreefall2}
\end{figure}

 The velocity distribution in a wave-packet, comprised of a coherent superposition of two gravitational states, includes several narrow peaks. These peaks are reproduced in the time-of-fall distribution. The positions of the two narrowest peaks, shown in Fig.\ref{FigFreefall2}, are equaal $t_1=246.9$ ms and $t_2=247.8$ ms, with each width equal to $0.5$ ms. The corresponding relative uncertainty is $\varepsilon\approx 2\cdot 10^{-3}$.

  Moreover, the relative positions of these peaks provide an access to the characteristic momentum distribution of a gravitational state via Eq.(\ref{pt}). Namely, the knowledge of values maxima in the time distribution of free-fall events $t_m$ could be related to the position of corresponding maxima in the momentum distribution $p_m=\hbar k_m/l_0$, where $k_m$ mean dimensionless values:
 \begin{equation}
 \frac{\hbar k_m}{l_0}=Mg(t_m-t_0).
 \end{equation}
 This fact gives another access to experimental measurements of a characteristic energy scale $\varepsilon_0=Mgl_0$:
 \begin{equation}
 \varepsilon_0=\frac{\hbar k_m}{t_m-t_0},
 \end{equation}
 and thus to the gravitational mass of antihydrogen:
 \begin{equation}
M=\sqrt{\frac{2m\hbar k_{m}^3 }{g^2(t_m-t_0)^3}}.
\end{equation}
With $10^3$ annihilation events this approach provides a relative accuracy of the order of $10^{-4}$ for the value of $M$.
 %%%%%%%%%%%%%%%%%%%%%%%%%%%%%%%%%%%%%%%%%%%%%%%%%%%%%%%%%%%%%%%%%%%%%%%%%%%
%%%%%%%%%%%%%%%%%%%%%%%%%%%%%%%%%%%%%%%%%%%%%%%%%%%%%%%%%%%%%%%%%%%%%%%%%%%
\section{Conclusion}
We propose a novel approach to study gravitational properties of antiatoms based on the spectroscopy of quantum states of $\bar{H}$ in the gravitational field of the Earth near a material surface. We showed that the gravitational mass $M$ of the $\bar{H}$ atom could be deduced from measuring the level spacing between gravitational states by means of resonance transitions, induced by an alternating gradient magnetic field. We study main properties of the interaction of gravitationally bound $\bar{H}$ with the magnetic field and calculate the probabilities of resonance transitions from the ground to excited states. In the proposed approach, the number of annihilation events is measured as a function of the applied field frequency. The spatial positioning of the annihilation detector at a height, corresponding to the classical turning point of the second gravitational state ensures that only $\bar{H}$ atoms in the final state are detected. The field gradient required for such a transition, $\beta=27.2$ Gs/m, as well as the guiding field $B_0=30$ Gs are easily achievable in experiments.

The width of the spectral line is determined by the time of life of gravitational states $\tau=0.1$ s. Using the transition from the ground to the fifth excited state and assuming that the number of detected annihilation events is $100$, the gravitational mass of $\bar{H}$ could be deduced with the precision $\sim 10^{-3}$. Though systematic effects are assumed to be small, their accurate estimation is underway.

A resonant magnetic field could be used to produce a coherent superposition of a few (at least two) gravitational states. The corresponding time-of-fall distribution in case of such a prepared wave-packet falling down from a given height manifests pronounced narrow peaks, which correspond to an easily predictable momentum distribution in the initial superposition. This method allows us to obtain the value of $M/m$ with an improved accuracy.
%\bibliographystyle{unsrt}
%\bibliographystyle{prsty}
%\bibliography{hbarclock}

\end{document}